\documentstyle[12pt]{article}
\begin{document}

Lietuvos fizikos \v zurnalas, 1998, {\bf 38}, Nr. 1, 118-121

Lithuanian Journal of Physics, 1998, {\bf 38}, Nr. 1, 118-121

\vspace{1.2cm}

\begin{center}
{\large {\bf QUANTUM ZENO AND QUANTUM ANTI-ZENO EFFECTS }}

\vspace{0.5cm}

{\large V. Gontis and B. Kaulakys}$^{*}$

\vspace{0.5cm}

{\it Institute of Theoretical Physics and Astronomy, A. Go\v stauo 12, 2600
Vilnius, Lithuania }$^{*}${\it and Vilnius University, Department of Quantum
Electronics }\vspace{0.5cm}

\parbox{5in}{\small Consequences of the deviation from the linear on time
quantum transition probabilities leading to the non-exponential decay law
and to the so-called Zeno effect are analysed. Main features of the quantum
Zeno and quantum anti-Zeno effects for induced transitions are revealed on
simple model systems.}

\vspace{0.5cm}
\end{center}

\section{Introduction}

It is generally accepted that unstable systems decay according to an
exponential law which has been experimentally verified on many quantum
systems. However, deviation from exponential decay have been predicted for
short as well as for very long decay times. The short-time deviation turns
out to be very interesting due to its consequences leading to the so-called
quantum Zeno effect -- the inhibition of decay by repeated observation in
very early stages of evolution. Here we discuss shortly the essence of the
exponential and non-exponential (nonlinear) decay and effect of frequent
measurements on the induced dynamics of simple and multilevel quantum
systems.

\section{Exponential and non-exponential decay}

When the decay probability $P_d$ of the system depends linearly on time, $
P_d=\gamma \tau $, for very short time intervals $\tau $ soon after
preparation of the system, $\gamma \tau \ll 1$, then the survival
probability of the system in the initial state is
$$
P\left( t\right) =1-\gamma \tau ,\quad \gamma \tau \ll 1.
$$
The survival probability as a result of evolution of the system during time $
t=n\tau $ is
$$
P\left( t=n\tau \right) =\left( 1-\gamma \tau \right) \left( 1-\gamma \tau
\right) \cdots \left( 1-\gamma \tau \right) =\left( 1-\gamma \tau \right)
^n. \eqno{(1)}
$$
For the large number of short duration evolution intervals, $n\rightarrow
\infty ,\tau \rightarrow 0,t=n\tau =const$, from Eq. (1) we have the
exponential law
$$
P\left( t\right) =\left( 1-\frac{\gamma n\tau }n\right) ^n=\left( 1-\frac{
\gamma t}n\right) ^n=\exp \left( -\gamma t\right) ,\left( n\rightarrow
\infty \right) .\eqno{(2)}
$$
Therefore, the linear in time decay process results in the exponential
survival probability dependence on the time. Exponential decay is a well
known property of the systems whose decay rate is proportional to its
undecayed quantity.

On the contrary, if a quantum system undergoes relatively slow, quadratic,
transitions to another states soon after preparation or measurement, $
P_d=g\tau ^2$, then survival in the initial state probability after time $
t=n\tau $ of the system subjected to $n-1$ intermediate measurements at
intervals $\tau $ is
$$
\begin{array}{c}
P\left( t=n\tau \right) =\left( 1-g\tau ^2\right) ^n=\left( 1-\frac{gn^2\tau
^2}{n^2}\right) ^n=\left( 1-\frac{\left( gt^2/n\right) }n\right) ^n
\end{array}
$$
$$
\simeq \exp \left( -gt^2/n\right) \rightarrow 1,\quad \left( n\rightarrow
\infty ,\tau \rightarrow 0,t=n\tau =const\right) \eqno{(3)}
$$
Therefore, as a result of such evolution with large number of intermediate
measurements the quantum system under consideration remains in the initial
state.

In the axiomatics of quantum mechanics it is postulated that any measurement
of the quantum system's state projects it onto an eigenstate of the measured
observable and causes disappearance of coherence of the system's state. For
this reason in the case of quadratic evolution in time of the system soon
after preparation or measurement, the repetitive frequent observation of the
quantum system can inhibit the decay of unstable [1] system and suppress
dynamics of the driven by an external field [2, 3] system. This phenomenon,
namely the inhibition or even prevention of the time evolution of the system
from an eigenstate of observable into a superposition of eigenstates by
repeated frequent measurements during the time of non-exponential dynamics,
is called the {\it quantum Zeno effect} (paradox) [1--4].

Theoretical analysis of deviation from exponential decay has a long history
[5]. It is possible to show quite generally from the equations of quantum
mechanics that the decay is slower than exponential for both very short and
very long times. The proof of such deviations requires only two very general
properties of the system (see [5, 6] and references therein): existence of
the lower bound to the energy that the decay products can have and the
decaying state must have finite energy. In such a case even from the simpler
perturbation theory for short evolution time $t\ll \hbar /\left(
E_u-E_l\right) $ we have
$$
P_d=\frac 1{\hbar ^2}\int\limits_{E_l}^{E_u}\left| V_{EE_0}\right| ^2\frac{
\sin ^2\frac{E-E_0}{2\hbar }t}{\left( \frac{E-E_0}{2\hbar }\right) ^2}\rho
\left( E\right) dE
$$
$$
\simeq \frac{E_u-E_l}{\hbar ^2}\left| V_{EE_0}\right| ^2\rho \left(
E_0\right) t^2\eqno{(4)}
$$
$$
P=1-P_d=1-gt^2\eqno{(5)}
$$
Here $E_u$ and $E_l$ are the upper and lower possible energy of the decaying
system, respectively, $V_{EE_0}$ is the matrix element of the interaction
potential for transition between the initial state with energy $E_0$ and the
products' state with energy $E$ and $\rho \left( E\right) $ is the density
of states of the decay products. These results are independent on detailed
interaction that causes the decay.

Usually the time-scales of the short-times deviation from the exponential
law are very small: about $10^{-23}$s for nuclei, $10^{-17}$s for atoms ($
3.6\times 10^{-15}$s for 2P-2S transition of the hydrogen atom [7]). For
such systems, therefore, we have no chance to observe the effect under
consideration. However, recently [8] experimental evidence for short-time
deviation from exponential decay in quantum tunnelling of ultracold sodium
atoms trapped in accelerating optical potential has been presented at the
microsecond time-scale.

The quantum Zeno effect, however, is easier tractable and observable not for
spontaneous decay to the continuum states but for the induced transitions
between discrete states of the quantum system [2, 3, 9] where transition
probabilities are nonlinear functions of time for relatively long time
intervals. Therefore, further we will analyse only transitions induced by
external field.

\section{Quantum Zeno effect for induced transitions}

Modeling the effect of measurement on the dynamics of quantum system by
randomizing the phases of the measured states we may derive equation for the
transition probabilities of the two-state system in the resonance field. The
simplest time evolution of the two-state wave function $\Psi =a_1\left|
1\right\rangle +a_2\left| 2\right\rangle $ from time moment $t_k=k\tau $ to $
t_{k+1}=(k+1)\tau $ can be represented as
$$
\left( \matrix{a_1(k+1)\cr a_2(k+1)}\right) ={\bf A}\left(
\matrix{a_1(k)\cr a_2(k)}\right) ,\eqno(6)
$$
$$
{\bf A}=\left(
\matrix{\cos\varphi& i\sin\varphi\cr
i\sin\varphi & \cos\varphi}\right) ,~~~\varphi ={\frac 12}\Omega \tau
\eqno(7)
$$
where $\Omega $ is the Rabi frequency. Evidently the evolution of the
amplitudes from time $t=0$ till $t=n\tau $ may be expressed as
$$
\left( \matrix{a_1(n)\cr a_2(n)}\right) ={\bf A}^n\left(
\matrix{a_1(0)\cr
a_2(0)}\right) .\eqno(8)
$$
One can calculate matrix ${\bf A}^n$ by the method of diagonalization of the
matrix ${\bf A.}$ The result naturally is
$$
{\bf A}^n=\left(
\matrix{\cos n\varphi & i\sin n\varphi\cr
i\sin n\varphi & \cos n\varphi}\right) .\eqno(9)
$$
For the time interval $T=n\tau =\pi /\Omega $ a certain (with the
probability $1$) transition between the states takes place.

Measurement of the system's state in the time moment $t=k\tau $ projects the
system to the state $\left| 1\right\rangle $ with the probability $
p_1(k)=\mid a_1(k)\mid ^2$ or to the state $\left| 2\right\rangle $ with the
probability $p_2(k)=\mid a_2(k)\mid ^2$. After each of the measurement the
phases of the amplitudes $a_1(k)$ and $a_2(k)$ are random which results in
the absence of the interference terms in the expressions for the
probabilities [9]. This results in the equation for the probabilities
$$
\left( \matrix{p_1(k+1)\cr p_2(k+1)}\right) ={\bf M}\left(
\matrix{p_1(k)\cr
p_2(k)}\right) ,\eqno(10)
$$
where
$$
{\bf M}=\left(
\matrix{\cos^2\varphi & \sin^2\varphi\cr
\sin ^2\varphi & \cos ^2\varphi}\right) \eqno(11)
$$
is the evolution matrix for the probabilities. The evolution from time $t=0$
until $t=n\tau $ with the $(n-1)$ intermediate measurement is described by
the equation
$$
\left( \matrix{p_1(n)\cr p_2(n)}\right) ={\bf M}^n\left(
\matrix{p_1(0)\cr
p_2(0)}\right) .\eqno(12)
$$
Matrix ${\bf M}^n$ calculated by the diagonalization method is
$$
{\bf M}^n={\frac 12}\left(
\matrix{1+\cos ^n2\varphi & 1-\cos ^n2\varphi\cr
1-\cos ^n2\varphi & 1+\cos ^n2\varphi}\right) .\eqno(13)
$$
From Eqs. (12) and (13) we get the quantum Zeno effect [2, 3]: if initially
the system is in state $\left| 1\right\rangle $, the outcome of evolution
until the time $T=n\tau =\pi /\Omega $ with the intermediate measurements
will be given by the probabilities $p_1(T)=(1+\cos ^n2\varphi )/2\rightarrow
1$ and $p_2(T)=(1-\cos ^n2\varphi )/2\rightarrow 0,~~(n\rightarrow \infty )$
. This result represents the inhibition of the quantum dynamics by
measurements and confirms the proposition that act of measurement may be
expressed as randomization of the amplitudes' phases.

\section{Quantum anti-Zeno effect for dynamics of multilevel systems}

Consider now the measurement effect on quantum dynamics of sufficiently more
complex systems, i.e. on the induced transitions between states of
multilevel system with the quantum suppression of chaotic dynamics. It is
natural to expect that frequent measurements of suppressed system will
result in additional inhibition of its dynamics.

In general the Schr\"odinger equation for strongly driven multilevel systems
can not be solved analytically. However, the mapping form of quantum
equations of motion greatly facilitates investigation of stochasticity and
quantum -- classical correspondence for chaotic dynamics. From the
standpoint of an understanding of manifestation of the measurements for the
dynamics of the multilevel systems the region of large quantum numbers is of
greatest interest. The simplest system in which the dynamical chaos and
quantum localization of states may be observed is a system with one degree
of freedom described by the nonlinear Hamiltonian $H_0(I)$ and driven by the
periodic $V(\theta ,t)=k\cos \theta \sum\limits_j\delta (t-j\tau )$ kicks
[10, 11]. Here the convenient variables, angle $\theta $ and action $I$, are
introduced. Integrating the Schr\"odinger equation over the period $\tau $
we obtain a map [11]
$$
a_m(t_{j+1})=e^{-i\beta _m}\sum\limits_na_n(t_j)J_{m-n}(k),~~\beta
_m=H_0(m)\tau ,~~t_j=j\tau \eqno(14)
$$
for the amplitudes $a_m(t_j)$ before the appropriate $j$-kick in expansion
of the state function $\Psi (\theta ,t)$ through the eigenfuntions, $\varphi
_m=i^{-m}e^{im\theta }/\sqrt{2\pi }$, of the action $I=-i\frac \partial
{\partial \theta }.$ Here $J_m(k)$ is the Bessel function and the phase
factor $i^{-m}$ is introduced for maximal simplification of the map.

Quantum dynamics represented by map (14) with the non-linear Hamiltonian $
H_0(I)$ is similar to the classical one only for some finite time $
t<t^{*}\simeq \tau k^2/2$, after which it reveals an essential decrease of
the diffusion rate asymptotically resulting in the exponential localization
of the system's state with the localization length $\lambda \sim k^2/2$ [10,
11].

Each measurement of the system's state between $(j-1)$ and $j$ kicks
projects it onto one of the state $\varphi _m$ with the probability $
P_m(t_j)=\left| a_m(t_j)\right| ^2.$ After such a measurement the phase of
the amplitude $a_m(t_j)$ is random. Therefore, the influence of the
measurements for further dynamics of the system may be expressed as
replacement of the amplitudes $a_m(t_j)$ by the amplitudes $\exp \left[
i2\pi g_m(t_j)\right] a_m(t_j)$, where $g_m(t_j)$ is a random number in case
of measurement of the $\varphi _m$-state's population before the $j$ kick
and equals zero in absence of such a measurement. So, we may analyze the
influence on the dynamics of measurements performed after every kick, after
every $N$ kicks or of the measurements just of some states, e.g. only of the
initial state, and observe the reduction of the quantum localization effect
in a degree depending on the extent and frequency of the measurement [9]. In
the case of measurement of all states after every kick we have the
uncorrelated transitions between the states and diffusion-like motion with
the quantum diffusion coefficient in the $n$-space
$$
B(n)={\frac 1{2\tau }}\sum\limits_m(m-n)^2J_{m-n}^2(k)={\frac{k^2}{4\tau }}
\eqno(15)
$$
which coincides with the classical one. Therefore, the quantum evolution of
frequently observable chaotic system is more classical-like than dynamics of
the isolated system (see also [12]). On the other hand, the repetitive
measurement of the multilevel systems with quantum suppression of classical
chaos results in delocalization of the states superposition and acceleration
of the chaotic dynamics which is opposite to the quantum Zeno effect in
simple few-level systems. Since this effect is the reverse of the quantum
Zeno effect we have called this phenomenon the {\it quantum anti-Zeno effect}
[9].

\section{Conclusion}

The essence and consequences of the exponential and non-exponential
(nonlinear) decay and the effect of repetitive measurement on quantum
dynamics of driven by an intensive external force of simple few-level
systems as well as of multilevel systems that exhibit the quantum
localization of classical chaos has been discussed. Frequent measurement of
the simple system yields to the quantum Zeno effect -- prevention of time
evolution, while that of the suppressed multilevel quantum system, which
classical counterpart exhibits chaos, results in the delocalization of the
quantum suppression. This outcome is the opposite to the quantum Zeno
effect. Therefore, we may call this phenomenon the {\it quantum anti-Zeno
effect}. Furthermore, in the limit of the frequent full measurement or
unpredictable interaction with the environment the quantum dynamics of
multilevel quasiclassical systems approaches the classical motion.

\vspace{0.5cm}

{\bf References }

\vspace{0.5cm}

\begin{enumerate}
\item  {\bf B. Misra and E. C. G. Sudarshan}, J. Math. Phys., V. 18, p. 756
(1976); {\bf C. B. Chiu, E. C. G. Sudarshan, and B. Misra}, Phys. Rev. D, V.
16, p. 520 (1977).

\item  {\bf R. J. Cook}, Phys. Scr., V. T21, p. 49 (1988).

\item  {\bf W. M. Itano, D. J. Heinzen, J. J. Bollinger, and D. J. Wineland}
, Phys. Rev. A, V. 41, p. 2295 (1990).

\item  {\bf P. Knight}, Nature, V. 344, p. 493 (1990); {\bf A. Schenzle},
Contemp. Phys., V. 37, p. 303 (1996).

\item  {\bf L. A. Khalfin}, Zh. Eksp. Teor. Fiz., V. 33, p. 1371 (1958)
[Sov. Phys. JETP, V. 6, p. 1503 (1958)]; {\bf J. Swinger}, Ann. Phys., V. 9,
p. 169 (1960); {\bf L. Fonda, G. C. Ghirardi, and A. Rimini}, Rep. Prog.
Phys., V. 41, p. 587 (1978); {\bf G.-C. Cho, H. Kasari, and Y. Yamaguchi},
Prog. Theor. Phys., V. V. 90, p. 803 (1993).

\item  {\bf P. T. Greenland}, Nature, V. 387, p. 548 (1997).

\item  {\bf P. Facchi and S. Pascazio}, Temporal behavior and quantum Zeno
region of an excited state of the hydrogen atom, {\it to be published}.

\item  {\bf S. R. Wilkinson at al.}, Nature, V. 387, p. 575 (1997).

\item  {\bf B. Kaulakys and V. Gontis}, Phys. Rev. A, V. 56, p. 1131 (1997);
e-print archives http://xxx.lanl.gov/quant-ph/9708024.

\item  {\bf G. Casati, B. V. Chirikov, D. L. Shepelyansky, and I. Guarneri},
Phys. Rep., V. 154, p. 77 (1987); {\bf F. M. Izrailev}, Phys. Rep., V. 196,
p. 299 (1990).

\item  {\bf V. G. Gontis and B. P. Kaulakys}, Liet. Fiz. Rink., V. 28, p.
671 (1988) [Sov. Phys.-Collec., V. 28(6), p. 1 (1988)].

\item  {\bf B. Kaulakys}, Lithuanian J. Phys., V. 36, 343 (1996); e-print
archives http://xxx.lanl.gov/quant-ph/9610041.

\end{enumerate}

\end{document}